\RequirePackage{fix-cm}
\documentclass[9pt]{extarticle}
\usepackage{eleco}  
\usepackage{subcaption} 

\usepackage{graphicx}
\usepackage{hyperref}
\usepackage{array} 
\usepackage{multirow}
\hypersetup{hidelinks,
  colorlinks=true,
  allcolors=black,
  pdfstartview=Fit,
  breaklinks=true}

\usepackage{array}
\newcommand{\PreserveBackslash}[1]{\let\temp=\\#1\let\\=\temp}
\newcolumntype{C}[1]{>{\PreserveBackslash\centering}p{#1}}

\usepackage{threeparttable}

\usepackage[capitalise,nameinlink]{cleveref}
\crefformat{equation}{#2(#1)#3}
\Crefformat{equation}{#2Equation (#1)#3}
\crefname{figure}{Fig.}{Figs.}

\title{Face Liveness Detection Using RGB and Thermal Image Fusion}
\name{Merve Erşan\textsuperscript{1,2}, Melike Girgin\textsuperscript{1,3}, and Tayfun Akgül\textsuperscript{1,3}}
\address{\textsuperscript{1}Department of Electronics and Communication Engineering, Istanbul Technical University, Istanbul, 34469, Türkiye \\
\textsuperscript{2}Informatics and Information Security Research Center (BİLGEM), TÜBİTAK, HİSAR Lab., Kocaeli, 41400, Türkiye\\
\textsuperscript{3} Advanced Research on Intelligent Systems Laboratory (ITU ARIS LAB), Department of Electronics and Communication Engineering, Istanbul Technical UniversityIstanbul, 34469, Türkiye \\
ersan17@itu.edu.tr, girginm20@itu.edu.tr, tayfunakgul@itu.edu.tr}

\begin{document}
\bstctlcite{IEEEexample:BSTcontrol}

\maketitle

\begin{abstract}

Face detection with visible-spectrum cameras can capture facial features, but it often fails to distinguish live subjects from spoof sources such as photographs, masks, or statues. Previous approaches based on texture, motion, or physiological cues are sensitive to illumination changes and show limited robustness against spoofing attacks. Thermal imaging helps overcome these limitations by detecting heat emissions, naturally excluding spoof faces. This study proposes a hybrid approach that fuses the edge information of RGB images with corresponding thermal images using a custom ARISTOF dataset containing live and spoof faces. The fused images are first evaluated using the YOLOv8-Face model to compare face detection performance across RGB, thermal, and fused modalities. The results show that the proposed method enhances the face detection accuracy of thermal images. The fused images are subsequently used to train a YOLOv8-Face model for live and spoof classification, demonstrating that the proposed multimodal fusion effectively supports robust face liveness detection.

\end{abstract}

\section{Introduction}\label{sec1}

Face detection is a fundamental task in computer vision with applications in security, surveillance, human–computer interaction, and biometric authentication \cite{authBiom,security,surveyFace,survSec}. RGB images provide rich structural and textural information, enabling accurate detection of faces in diverse environments. Extensive research has focused on detecting faces in RGB images due to their widespread availability and high spatial resolution, which allows for detailed feature extraction. However, RGB-based face detection systems alone cannot reliably distinguish between live and spoof faces, such as photographs, masks, or digital reproductions, which limits their effectiveness in face liveness detection tasks \cite{RGB-liveness}. Various approaches have been proposed to address this challenge, including the use of texture, motion, or physiological cues, yet these methods often remain sensitive to changes in lighting, occlusions, and spoofing attempts \cite{RGB-livEx1,RGB-livEx2,RGB-livEx3,RGB-livEx4,RGB-livEx5}.

Thermal imaging offers complementary advantages by capturing the heat emission patterns of living subjects. Since non-living objects do not emit heat in the same manner, thermal cameras inherently reduce the risk of misclassifying spoof faces. Additionally, thermal imaging is robust under varying illumination conditions, providing reliable liveness information even in low-light or dark environments \cite{thInf1,thInf2,thInf3}. Despite these benefits, thermal-only systems often lack the high spatial resolution and detailed structural information available in RGB images, which can limit precise face detection \cite{rgb-th1,rgb-th2}.

Motivated by the complementary strengths of RGB and thermal modalities, this study introduces ARISTOF, a custom hybrid dataset comprising paired RGB–Thermal (RGB-T) images of live and spoof faces. In order to create hybrid representations, edges are extracted from RGB images using Sobel operator and fused with corresponding thermal data. RGB, thermal, and fused inputs are initially examined using the YOLOv8-Face model to compare face detection performance among these modalities. Results indicate that incorporating edge-based fusion enhances face detection accuracy in thermal images. Subsequently, the fused images are employed to train a YOLOv8-Face model for live and spoof classification. Experimental results indicate that this multimodal approach enables reliable face liveness detection, highlighting the benefits of combining RGB and thermal information.

The main contributions of this study can be summarized as:

\begin{itemize}
    \item A new dataset of paired RGB and thermal images, containing both live and spoof faces, named ARIS LAB Thermal Optic Faces (ARISTOF), is constructed.
    \item Face detection performance on thermal images is improved by fusing edges extracted from RGB images with their corresponding thermal images.
    \item A YOLOv8-based model is trained on the annotated live and spoof faces, and its effectiveness in liveness detection is evaluated.
\end{itemize}

The remainder of this paper is organized as follows. Section \ref{sec2} describes the collection and preparation of the new ARISTOF dataset. Section \ref{sect3} introduces the proposed method. Section \ref{sec4} presents implementation details and experimental results. Finally, Section \ref{conclusion} concludes the paper.

\section{Dataset Collection and Preparation}\label{sec2}

During the generation of the ARISTOF dataset, thermal and RGB images are collected using a FLIR One Pro LT, a hybrid camera featuring co-aligned visible-spectrum and thermal sensors, mounted on an iPhone 14 Pro. The visible camera has a resolution of $1440\times1080$ pixels, while the thermal camera has a resolution of $80\times60$ pixels, covering a scene temperature range from $-20^\circ$C to $120^\circ$C \cite{FLIROneProLT2025}.

\begin{figure}[h]
    \centering
    \includegraphics[width=0.48\textwidth]{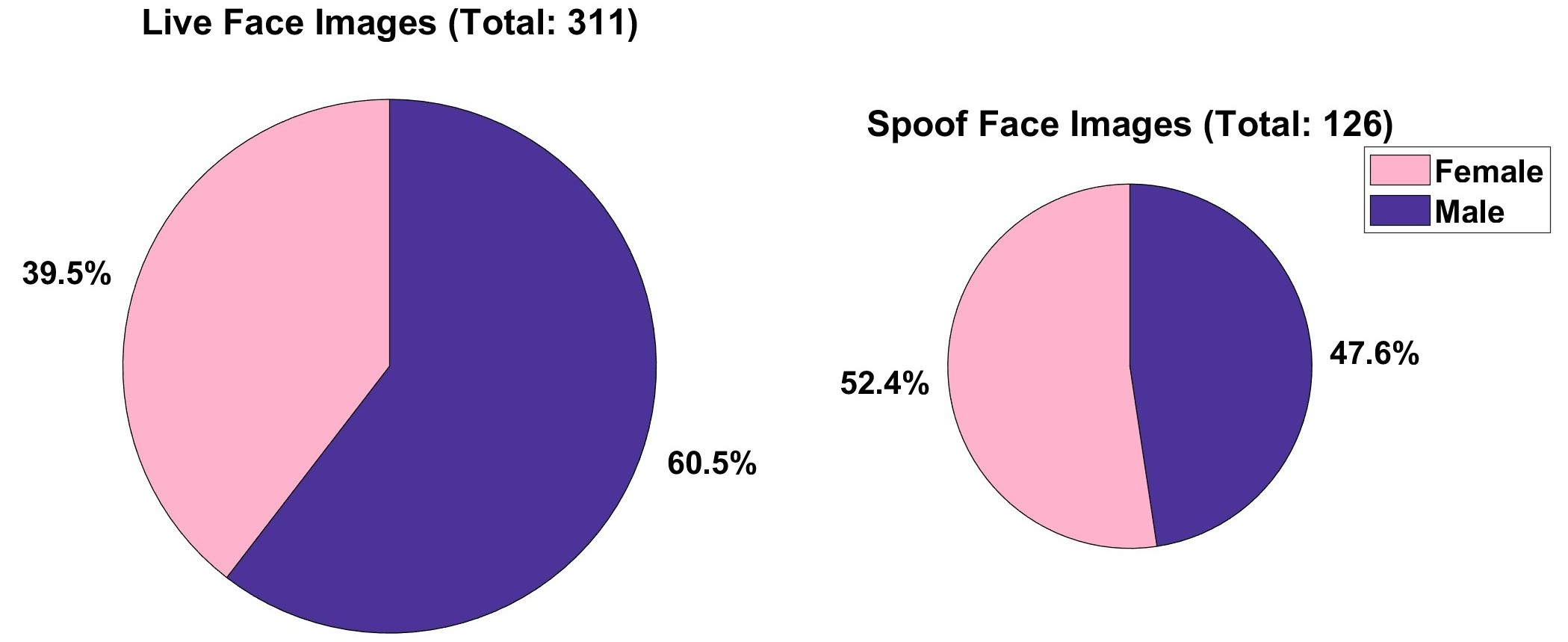}
    \caption{Gender distribution of live and spoof faces. Each pie chart shows the proportion of Female (pink) and Male (purple) images for Live (left) and Spoof (right) categories.}
    \label{fig:donut_distribution}
\end{figure}

The dataset contains 311 images of live human faces and 126 spoof faces which are the photographs of human faces. Among the live faces, 34 belong to female subjects and 49 to male subjects, while the spoof faces include 41 females and 33 males. The gender distribution of live and spoof images is illustrated in Fig.~\ref{fig:donut_distribution} using pie charts.

\begin{figure*}[h!]
	\centering	
	\includegraphics[width=0.85\textwidth ]{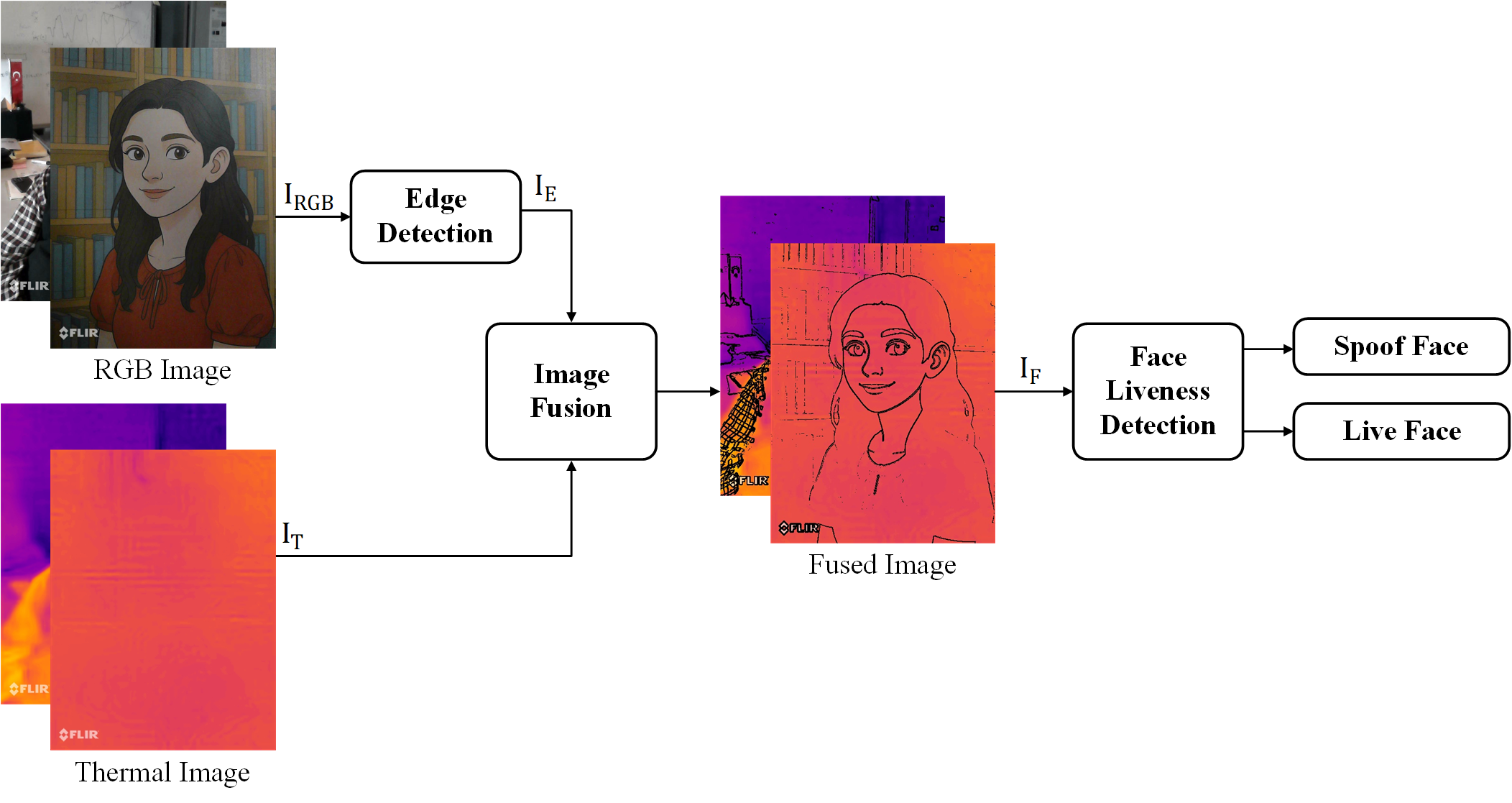}	
	\caption{System model: Edge information ($I_{E}$) is extracted from the RGB images ($I_{RGB}$) and fused with the corresponding thermal images ($I_{T}$) to obtain complementary information from both modalities. The resulting fused images ($I_{F}$) are then annotated as live or spoof and used for face liveness detection.}
	\label{systemmodel}
\end{figure*}

Portraits of subjects are taken from multiple angles, such that each face occupies approximately \%10 of the image area. To enhance dataset diversity and robustness, some subjects wear accessories such as glasses or hats. Moreover, data are collected under both well-lit and low-light conditions, providing a wide range of visual variations for training and evaluation.

All participants were informed and gave their consent prior to image acquisition for research purposes only. However, due to privacy considerations and KVKK (Personal Data Protection Law in Türkiye) regulations, the dataset is not publicly available.

\section{Proposed Method} \label{sect3}

This section presents the proposed method for enhancing face liveness detection using hybrid RGB–T imaging. The method extracts edges from RGB images and fuses them with the corresponding thermal images to generate hybrid RGB–T data, which are subsequently used for face liveness detection. The overall system model of the proposed approach is illustrated in Fig.~\ref{systemmodel}.

\subsection{Edge Detection and Image Fusion} \label{imageFusion}

The ARISTOF dataset is collected using a co-aligned hybrid camera, which eliminates synchronization issues between the thermal and visible-spectrum modalities. While stereo calibration is generally required when hybrid images include objects at varying distances, in this study, faces are captured at close range, occupying approximately \%10 of the image area. Consequently, a simple vertical (y-axis) shift based on camera positioning is sufficient to ensure spatial alignment between the RGB and thermal images. 

The first step of our proposed method involves edge detection of RGB images. To identify the most suitable method for our application, four commonly used edge detection techniques, Laplacian of Gaussian (LoG) \cite{tomasi1998bilateral}, Prewitt \cite{prewitt1970object}, Sobel \cite{sobel19683x3}, and Canny \cite{canny2009computational}, are applied to sample RGB images from the ARISTOF dataset. The resulting edge maps are then fused with the corresponding thermal images according to
\begin{equation} \label{eq:fusion}
I_{\text{F}}(x,y) = \alpha I_{T}(x,y) + (1-\alpha) I_E(x,y),
\quad 0 < \alpha < 1,
\end{equation}
where $I_{T}(x,y)$ denotes the thermal image, $I_{E}(x,y)$ represents the edge map of the RGB image, and $\alpha$ is a weighting factor that determines the relative contribution of the thermal image compared to the edge map. In our experiments, $\alpha$ is set to 0.5. Consequently, the fused image $I_{\text{F}}(x,y)$ is generated by combining the edge-extracted RGB image with its corresponding thermal image.

\begin{figure*}[h!]
    \centering

    \begin{minipage}{0.18\textwidth}
        \centering
        \includegraphics[width=\textwidth]{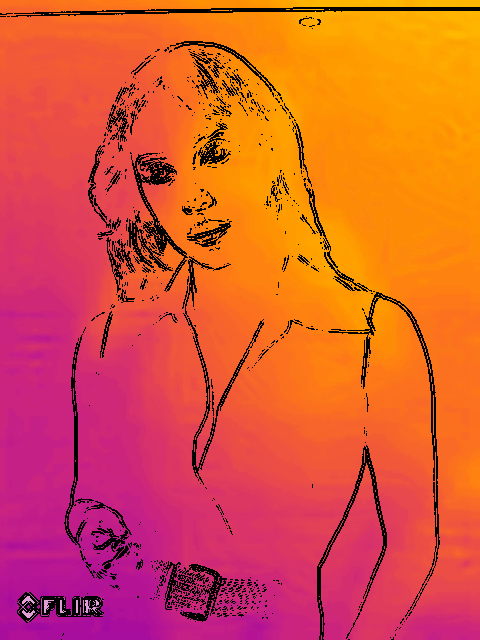}
        \vspace{2pt}
        {\small (a)}
    \end{minipage}\hspace{0.015\textwidth}%
    \begin{minipage}{0.18\textwidth}
        \centering
        \includegraphics[width=\textwidth]{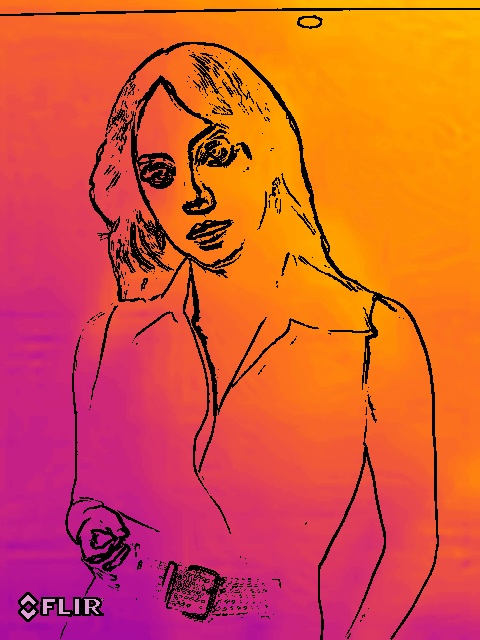}
        \vspace{2pt}
        {\small (b)}
    \end{minipage}\hspace{0.015\textwidth}%
    \begin{minipage}{0.18\textwidth}
        \centering
        \includegraphics[width=\textwidth]{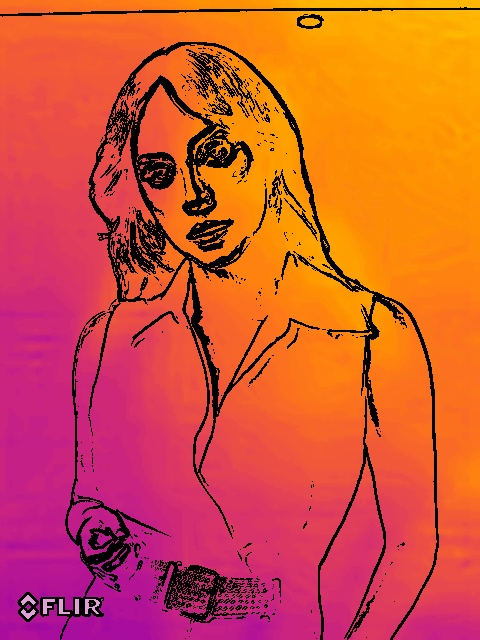}
        \vspace{2pt}
        {\small (c)}
    \end{minipage}\hspace{0.015\textwidth}%
    \begin{minipage}{0.18\textwidth}
        \centering
        \includegraphics[width=\textwidth]{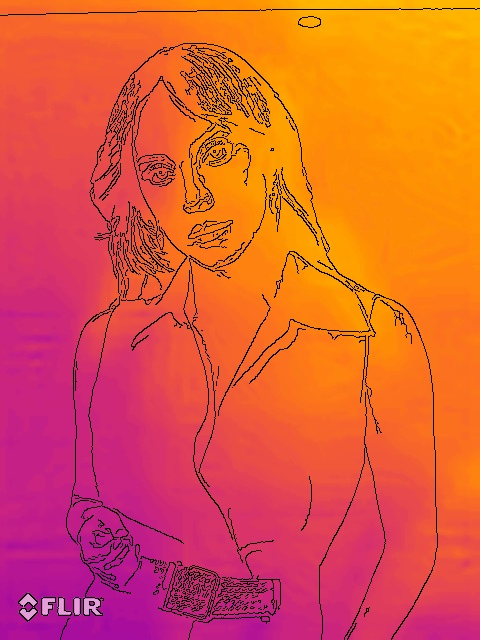}
        \vspace{2pt}
        {\small (d)}
    \end{minipage}

    \caption{Fused RGB–T images generated using four edge detection methods:  
    (a) Laplacian of Gaussian (LoG), (b) Prewitt, (c) Sobel, and (d) Canny.}
    \label{fig:edge_comparison}
\end{figure*}

\cref{fig:edge_comparison} illustrates the fused images obtained by applying the four edge detection techniques to ARISTOF dataset samples, according to Equation~\ref{eq:fusion}. The figure enables visual inspection of the fusion outcomes and highlights how different edge detection algorithms influence the resulting hybrid images. Among the evaluated methods, Sobel is selected as it preserves facial details more effectively than the alternatives.

In Sobel, the horizontal kernel $g_x$ and vertical kernel $g_y$ are defined as
\begin{equation}
    g_x = \begin{bmatrix}
    -1 & 0 & +1 \\
    -2 & 0 & +2 \\
    -1 & 0 & +1
    \end{bmatrix}, \ \ \
    g_y = \begin{bmatrix}
    -1 & -2 & -1 \\
     0 &  0 &  0 \\
    +1 & +2 & +1
    \end{bmatrix}.
\end{equation}
The horizontal gradient
\begin{equation}
    \begin{aligned}
        G_x(x,y) &= \sum_{i=-1}^{1} \sum_{j=-1}^{1} g_x(i,j) \, I_{RGB}(x+i,\, y+j) \\
    \end{aligned}
\end{equation}
and vertical gradient
\begin{equation}
    \begin{aligned}
        G_y(x,y) &= \sum_{i=-1}^{1} \sum_{j=-1}^{1} g_y(i,j) \, I_{RGB}(x+i,\, y+j)
    \end{aligned}
\end{equation}
are computed by convolving the kernels with the RGB image (\(I_{RGB}\)), respectively. Then, the edge image, \( I_{E}(x,y) \), obtained from the gradient magnitude calculation is 
\begin{equation}
    I_E(x,y) = \sqrt{G_x(x,y)^2 + G_y(x,y)^2}.
\end{equation}

\begin{figure}[h!]
    \centering

    \begin{subfigure}[b]{0.32\linewidth}
        \centering
        \includegraphics[width=\linewidth]{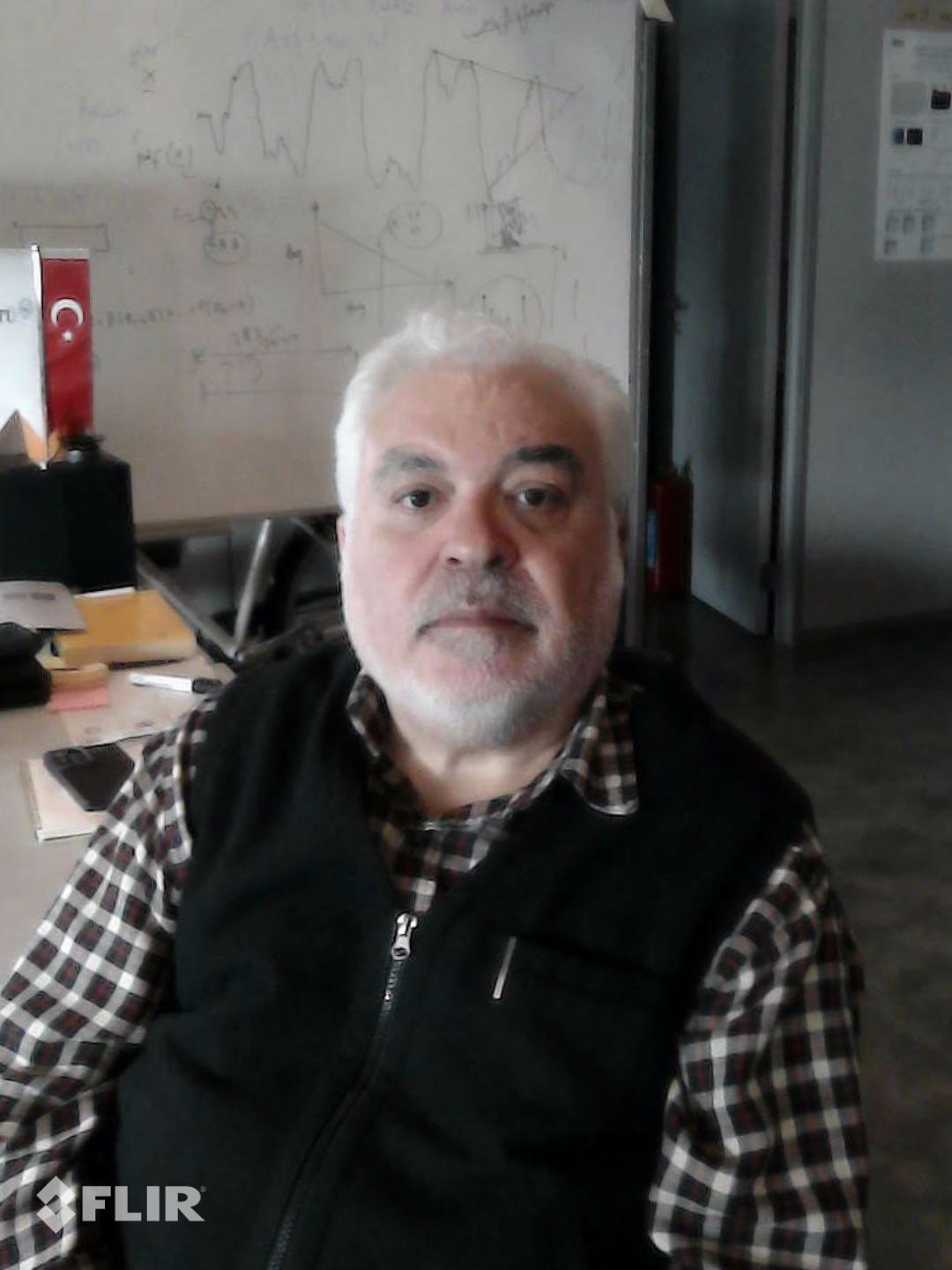}
        \caption{}
        \label{fig:sub_real_rgb}
    \end{subfigure}
    \hfill
    \begin{subfigure}[b]{0.32\linewidth}
        \centering
        \includegraphics[width=\linewidth]{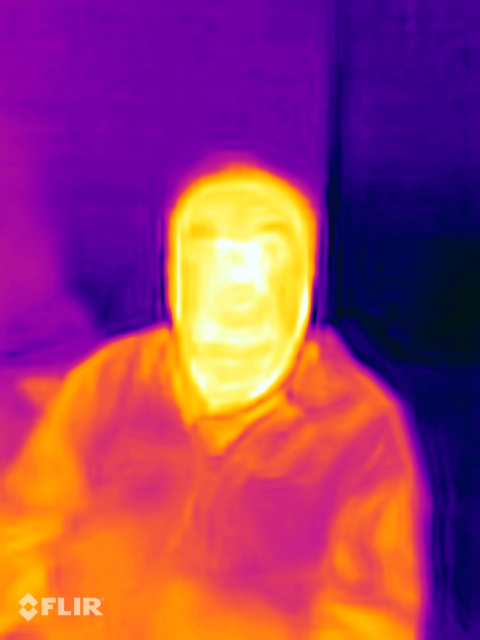}
        \caption{}
        \label{fig:sub_real_thermal}
    \end{subfigure}
    \hfill
    \begin{subfigure}[b]{0.32\linewidth}
        \centering
        \includegraphics[width=\linewidth]{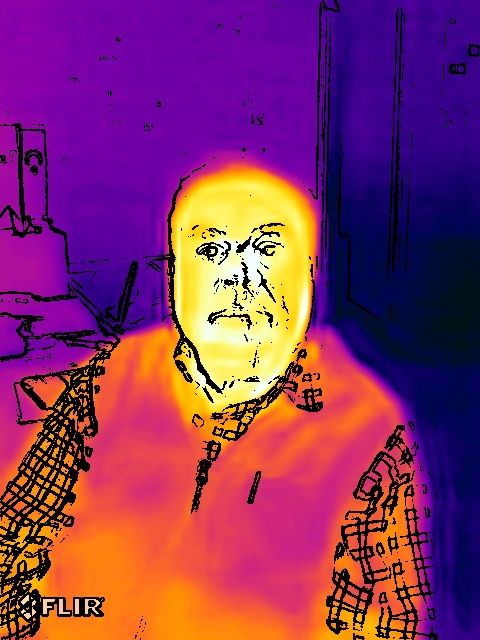}
        \caption{}
        \label{fig:hybrid_real}
    \end{subfigure}

    \vspace{2mm} 

    \begin{subfigure}[b]{0.32\linewidth}
        \centering
        \includegraphics[width=\linewidth]{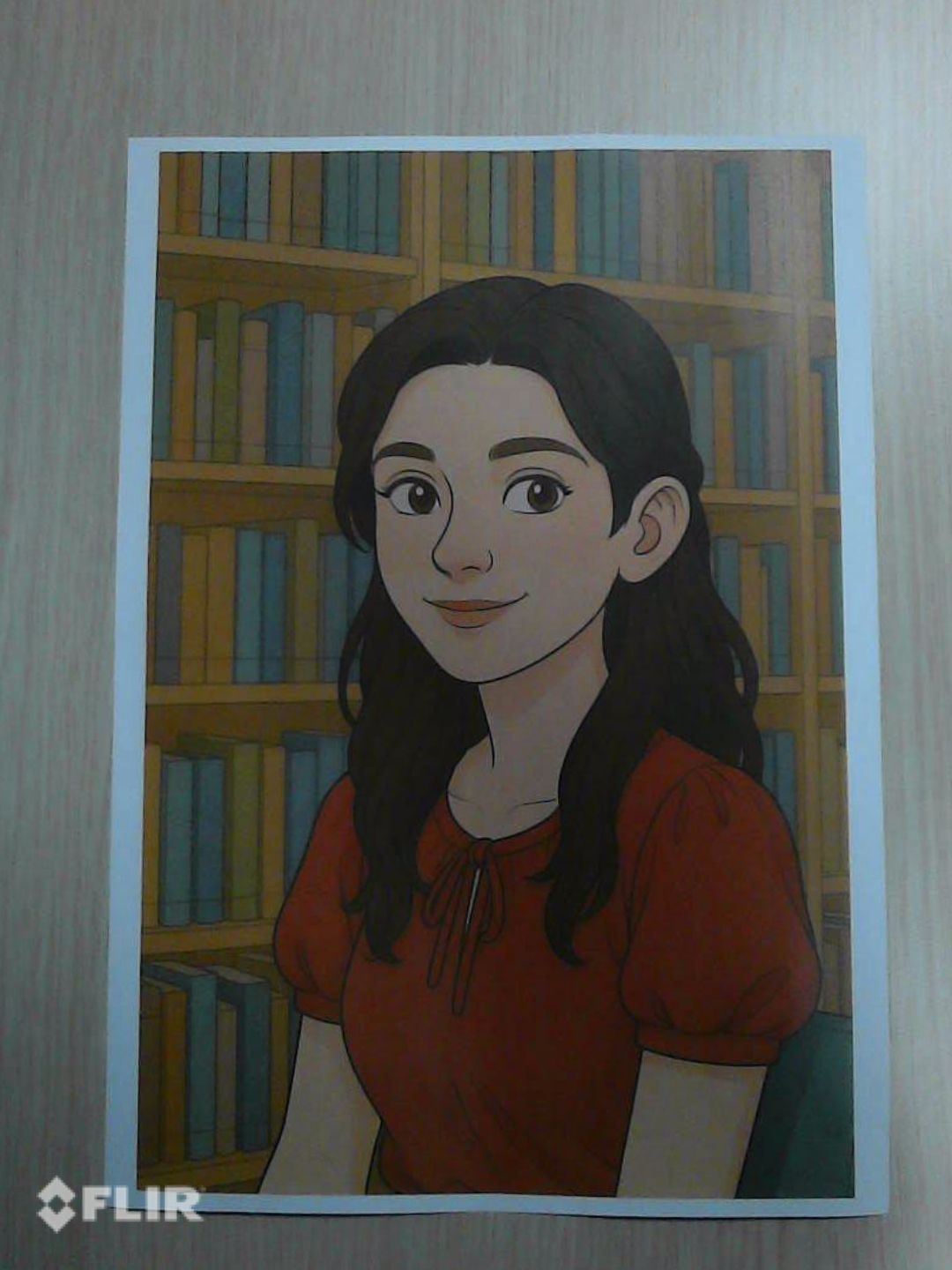}
        \caption{}
        \label{fig:sub_fake_rgb}
    \end{subfigure}
    \hfill
    \begin{subfigure}[b]{0.32\linewidth}
        \centering
        \includegraphics[width=\linewidth]{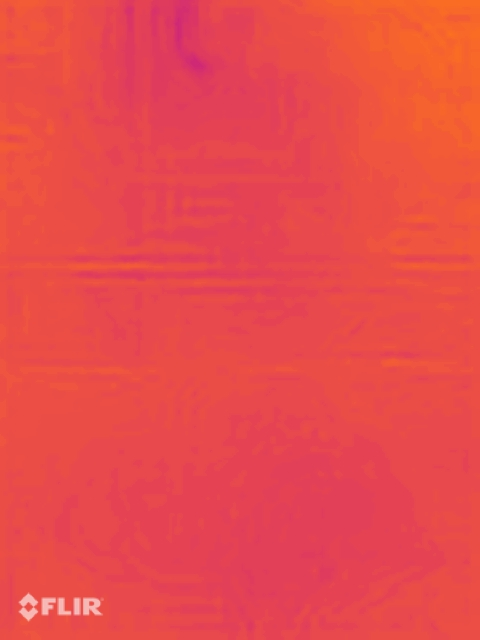}
        \caption{}
        \label{fig:sub_fake_thermal}
    \end{subfigure}
    \hfill
    \begin{subfigure}[b]{0.32\linewidth}
        \centering
        \includegraphics[width=\linewidth]{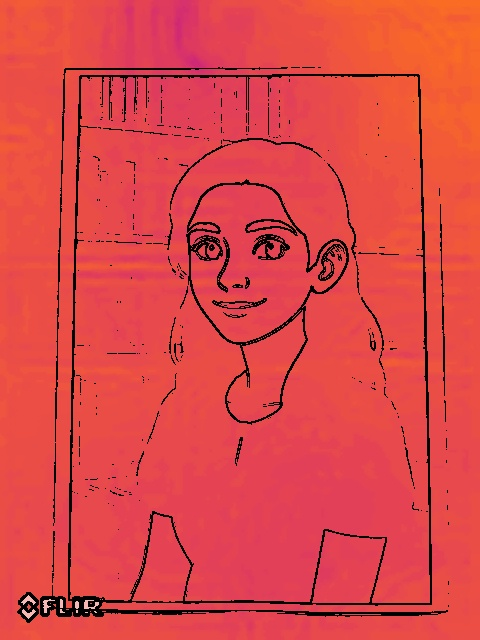}
        \caption{}
        \label{fig:hybrid_fake}
    \end{subfigure}

    \caption{The images shows the collected live and spoof face images and their fused versions: (a) live face in RGB; (b) live face in thermal; (c) fusion of (a) and (b); (d) spoof face in RGB, (e) spoof face in thermal, (f) fusion of (d) and (e).}
    \label{fig:all_examples_2x3_right}
\end{figure}

\noindent The fused image \( I_{\text{F}}(x,y) \) is obtained by combining the edge information from the RGB image \( I_{\text{E}}(x,y) \) with the corresponding thermal image \({I}_{T}(x,y) \) using
Equation~\ref{eq:fusion}. The hybrid results obtained using the proposed method are shown in Fig.~\ref{fig:all_examples_2x3_right}.

\vspace{5pt}

Face detectability in thermal images is noticeably enhanced by the proposed method, as clearly shown in the visual results.

\subsection{Face Liveness Detection}

The YOLOv8-Face model, a state-of-the-art real-time object detection framework optimized for face detection, is employed to evaluate the effectiveness of the hybrid RGB–T images. This model offers a favorable balance between detection accuracy and computational efficiency, making it suitable for training on custom datasets with diverse visual conditions. However, despite its strong face detection capability, YOLOv8-Face does not provide any liveness information and cannot distinguish between live and spoof faces.

After the hybrid RGB–T images are generated using the proposed method, they are uploaded to Roboflow, where each image is annotated as either “live face” or “spoof face". The dataset is split into training, validation, and test sets with a ratio of 70\%, 15\%, and 15\%, respectively. Data augmentation techniques, including horizontal flipping, rotation, and shear, are applied to the training set to increase variability and improve model robustness. After augmentation, the training set size increases approximately threefold. All images are preprocessed to ensure consistent input dimensions for the YOLOv8-Face model. Table~\ref{table:dataset_split} summarizes the number of images in each subset.

This setup ensures that the YOLOv8-Face model can learn from a balanced and representative set of hybrid images, capturing the structural cues provided by edge-enhanced RGB information and thermal data. 

\begin{table}[h!]
\centering
\caption{Number of live and spoof face classes in training, validation, and test sets}
\label{table:dataset_split}
\setlength\tabcolsep{3mm}
\begin{tabular}{|c|c|c|c|}  
\hline
Subset & Live face & Spoof face & Total \\ \hline
Training        & 654            & 255            & 909            \\ \hline
Validation      & 46             & 19            & 65             \\ \hline
Test            & 47             & 18            & 65             \\ \hline
\end{tabular}
\end{table}

Metrics such as Precision, Recall, and mean Average Precision at 50\% Intersection over Union (mAP@50) are used to evaluate model performance. Precision is calculated as $\text{TP}/(\text{TP}+\text{FP})$, measuring the proportion of correctly identified objects among all predictions, where TP (True Positive) denotes the number of correctly detected objects and FP (False Positive) denotes the number of incorrectly detected objects. Recall is computed as $\text{TP}/(\text{TP}+\text{FN})$, indicating the proportion of actual objects that are correctly detected, where FN (False Negative) represents the number of objects that the model fails to detect. The mAP@50 metric averages the Average Precision (AP) values across all classes, where each AP is determined by the area under the precision-recall curve obtained from varying detection thresholds, considering a prediction correct if its Intersection over Union (IoU) with the ground truth is at least 0.5.

\section{Experimental Setup and Results} \label{sec4}

The implementation and training details of the model are presented and the results are analyzed using quantitative metrics.

\subsection{Implementation and Training Details}

Training and evaluation are conducted using the YOLOv8-Face object detection model. The training is carried out in a Google Colab environment utilizing the PyTorch framework with an NVIDIA Tesla T4 GPU (15,095 MiB). The model is trained for 50 epochs with a batch size of 16 and an initial learning rate of 0.001. Optimization is performed using the Adam optimizer.

\subsection{Quantitative Results and Discussion}

In this paper, two analyses are presented: (i) the face detection performance of the YOLOv8-Face model across RGB-only, thermal-only, and fused images, and (ii) the face liveness detection performance of the proposed model.

Initially, the RGB and thermal images from the ARISTOF dataset, along with the fused images generated using the proposed method, are evaluated using the YOLOv8-Face model for face detection. It should be noted that YOLOv8-Face performs only face detection and does not provide any liveness information. The results are shown in Table~\ref{table:detection_rates}. The live and spoof face classes are provided solely for the reader's reference and to facilitate a clear analysis of detection performance.

\begin{table}[!h]
\centering
\begin{threeparttable}
\caption{Face detection rate of the YOLOv8-Face model across different input modalities.}
\label{table:detection_rates}
\setlength\tabcolsep{2mm} 
\begin{tabular}{|>{\centering\arraybackslash}p{0.28\columnwidth}|
                >{\centering\arraybackslash}p{0.2\columnwidth}|
                >{\centering\arraybackslash}p{0.35\columnwidth}|}
\hline
{Input Modality} & {Class} & {Face detection rate} \\ \hline
\multirow{2}{*}{\centering RGB-only} & Live face & 100\% \\ \cline{2-3}
 & Spoof face & 100\% \\ \hline
\multirow{2}{*}{\centering Thermal-only} & Live face & 65.9\% \\ \cline{2-3}
 & Spoof face & 0\% \\ \hline
\multirow{2}{*}{\centering Fused} & Live face & 75.2\% \\ \cline{2-3}
 & Spoof face & 65.6\% \\ \hline
\end{tabular}
\end{threeparttable}
\end{table}

As expected, the YOLOv8-Face model detects faces in 100\% of the collected RGB images for both live and spoof face classes. In contrast, no face detection occurs for spoof faces in thermal images, since these faces are captured from two-dimensional sources and do not emit heat. This indicates that thermal inputs alone lack the spatial and textural details necessary for reliable face detection. Live faces, however, emit heat, enabling the model to detect some of them, achieving a detection rate of 65.9\%. When fused RGB–T images are used, the detection rate increases to 75.2\% for live faces, while for spoof faces, it rises substantially from 0\% to 65.6\%. This improvement is attributed to the complementary spatial and textural information provided by the RGB edges, which allows the fused images to enhance face detection performance for both live and spoof faces compared to thermal-only images.

After observing that the proposed method improves face detection of the thermal-only images, we train the YOLOv8-Face for face liveness detection using the ARISTOF dataset, which is previously annotated with live and spoof faces and split into training, validation, and test sets. Training results can be seen in Table~\ref{table:detection_rates2}.

\begin{table}[!h]
	\centering
	\begin{threeparttable}
    \caption{Performance metrics of the proposed RGB–Thermal fusion method.}
	\label{table:detection_rates2}
	\setlength\tabcolsep{1.5mm}
	\begin{tabular}{|C{18mm}|C{16mm}|C{16mm}|C{16mm}|}
	\hline
	   {Class} & {Precision} & {Recall} & {mAP@50} \\ \hline
	Live face & 0.99 & 1 & 0.99 \\ \hline
	Spoof face & 0.94 & 0.95 & 0.93 \\ \hline
	\end{tabular}
	\end{threeparttable}
\end{table}

The training results show that the model achieves high performance when trained on the fused RGB–Thermal images. For live faces, the model attains a Precision of 0.99, Recall of 1.00, and mAP@50 of 0.99, indicating nearly perfect detection. Spoof faces are also detected with high accuracy, with a Precision of 0.94, Recall of 0.95, and mAP@50 of 0.93. These results demonstrate that the proposed fusion method effectively guides the model to learn discriminative features for both live and spoof face classes, ensuring robust performance during training.

In addition to the training results, the model is evaluated on unseen 65 test images, achieving a Precision of 1.00, Recall of 0.95, and mAP@50 of 0.97. These results confirm that the proposed fusion method maintains high detection accuracy under real test conditions.

\begin{figure}[h!]
    \centering
    \includegraphics[width=0.5\textwidth]{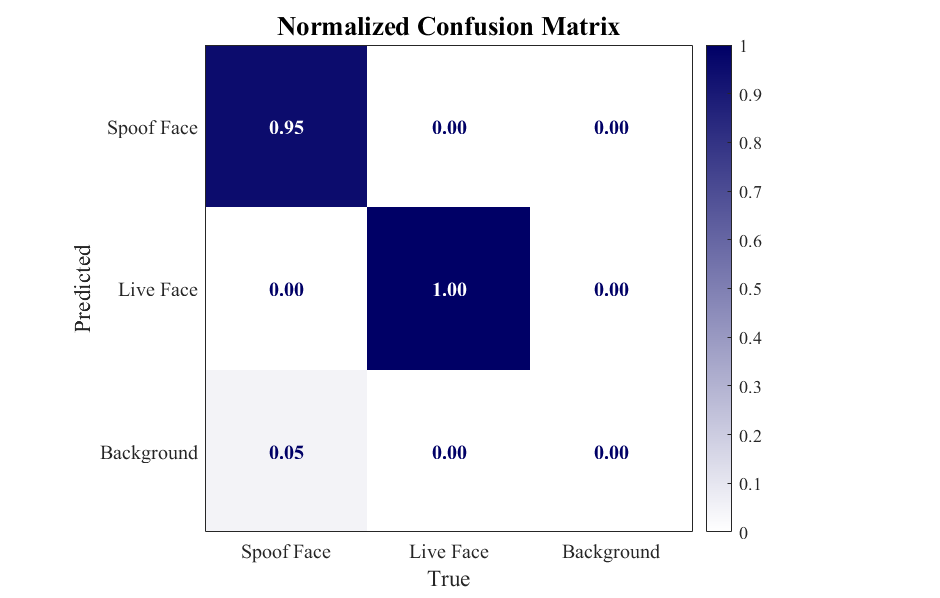}
    \caption{Normalized confusion matrix of the proposed hybrid RGB-Thermal fusion model on validation images.}
    \label{fig:confMx}
\end{figure}

The normalized confusion matrix of the proposed model is shown in Fig.~\ref{fig:confMx}. 
The results demonstrate that all live faces are correctly identified, with a true positive rate of 1.00, while 95\% of spoof faces are accurately classified. 
A small portion (5\%) of spoof faces is misclassified as background, indicating that some low-quality or partially occluded samples lack sufficient structural cues for reliable detection. 
Overall, the confusion matrix confirms the model’s high discriminative capability and consistency across both live and spoof classes.

\begin{figure}[h!]
    \centering
    \begin{subfigure}[b]{0.38\linewidth}
        \centering
        \includegraphics[width=\linewidth]{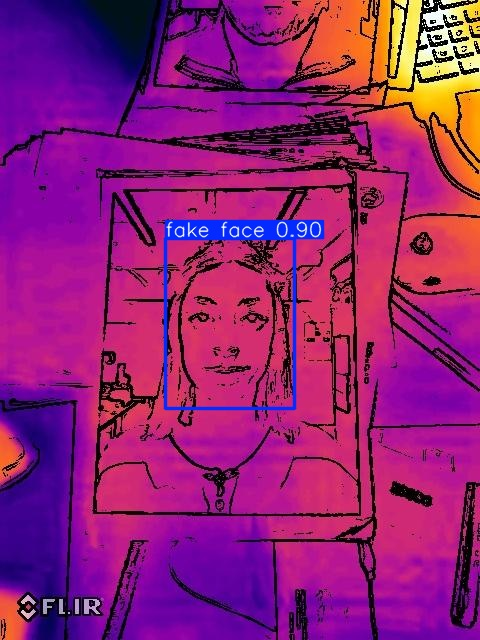}
    \end{subfigure}
    \hspace{2mm}
    \begin{subfigure}[b]{0.38\linewidth}
        \centering
        \includegraphics[width=\linewidth]{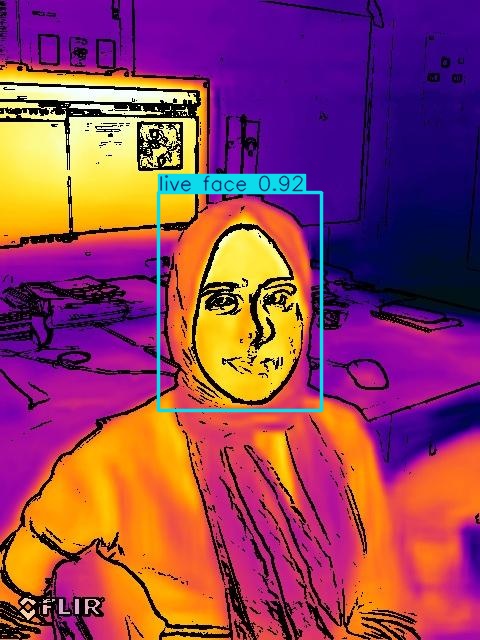}
    \end{subfigure}
    \caption{Example detections obtained by the proposed hybrid RGB-Thermal fusion method on test images, correctly detected (a) spoof face and (b) live face.}
    \label{fig:fused_faces}
\end{figure}

Figure~\ref{fig:fused_faces} illustrates example detections produced by the proposed model. 
In both cases, the model accurately distinguishes live and spoof faces with high confidence scores (0.90 and 0.92, respectively).

\section{Conclusions}\label{conclusion}

This work introduced a custom dataset of RGB–T image pairs containing both live and spoof faces, together with a hybrid fusion strategy in which edges extracted from RGB images were integrated with the corresponding thermal images. Face detection performance was initially evaluated using the YOLOv8-Face model on RGB, thermal, and fused inputs, showing that the proposed fusion method provides a clear improvement over thermal images alone. The fused images were then used to train a model for face liveness detection, achieving consistently satisfactory performance and demonstrating the effectiveness of the proposed hybrid RGB–T fusion approach for reliable liveness detection. For future work, the dataset could be extended to incorporate 3D spoofing artifacts along with dynamic sequences of live and spoof faces, enabling a more comprehensive evaluation of the model’s robustness.

\section{References}
\bibliography{IEEEabrv,biblio}
\end{document}